\documentclass[aip,jcp,twocolumn,10pt]{revtex4-1}
\usepackage{graphicx}%
\usepackage{dcolumn}
\usepackage{amsmath}   
\usepackage[usenames,dvipsnames]{pstricks}
\usepackage{epsfig}
\usepackage{pst-grad} 
\usepackage{pst-plot} 

\usepackage{graphicx} 
\usepackage{bm}
\usepackage{longtable}

\begin{document}

\title{ Non-Gaussian statistics of electrostatic fluctuations of hydration shells } 
\author{Allan D.\ Friesen}
\author{Dmitry V.\ Matyushov}\email{dmitrym@asu.edu}
\affiliation{Center for Biological Physics, Arizona State University,
  PO Box 871604, Tempe, AZ 85287-1604 } 

\begin{abstract}
  We report the statistics of electric field fluctuations produced by
  SPC/E water inside a Kihara solute given as a hard-sphere core with
  a Lennard-Jones layer at its surface. The statistics of electric
  field fluctuations, obtained from numerical simulations, are studied
  as a function of the magnitude of a point dipole placed close to the
  solute-water interface. The free energy surface as a function of the
  electric field projected on the dipole direction shows a cross-over
  with the increasing dipole magnitude. While it is a single-well
  harmonic function at low dipole values, it becomes a double-well
  surface at intermediate dipole moment magnitudes, transforming to a
  single-well surface, with a non-zero minimum position, at still
  higher dipoles.  A broad intermediate region where the interfacial
  waters fluctuate between the two minima is characterized by intense
  field fluctuations, with non-Gaussian statistics and the variance
  far exceeding the linear-response expectations. The excited state of
  the surface water is found to be lifted above the ground state by
  the energy required to break approximately two hydrogen bonds. This
  state is pulled down in energy by the external electric field of the
  solute dipole, making it readily accessible to thermal
  excitations. The excited state is a localized surface defect in the
  hydrogen-bond network creating a stress in the nearby network, but
  otherwise relatively localized in the region closest to the solute
  dipole.
\end{abstract}

\maketitle

\section{Introduction}
\label{sec:1}
Polarization of polar liquids by external electric fields is one of
classical problems of condensed matter theory.\cite{Landau8,Frohlich}
The ideas advanced by Debye and Langevine assign the average dipole
moment in a liquid exposed to an external electric field to a function
increasing monotonically and linearly (Debye) or non-linearly, with
saturation (Langevine) with the increasing field
strength.\cite{Boettcher:73} The polarized medium, in return, produces 
a field of its own which, combined with the external field, yields the
macroscopic Maxwell field. One thus arrives at the concept of
dielectric screening implying the ability, quantified by the 
macroscopic dielectric constant, of the medium to polarize
and produce its own electric field, in response to an external
perturbation.
 
The electric field of a polar liquid can be alternatively probed by an
external multipole. The electric field $\mathbf{R}(\Gamma)$ inside a
liquid is a fluctuating variable depending on the instantaneous liquid
configuration represented by a point in its phase space $\Gamma$.
Thus an external dipole moment $\mathbf{m}_0$ placed outside or inside
the liquid will gain the instantaneous electric energy
$e=-\mathbf{m}_0 \cdot \mathbf{R}$. The liquid, in turn, is polarized
by $\mathbf{m}_0$ such that its field at the position of the dipole
fluctuates around the average field $\mathbf{R}_0$, which is aligned 
along $\mathbf{m}_0$ and is known as the Onsager reaction
field.\cite{Onsager:36} This average field is accessible
experimentally from solvent-induced shifts of optical
dyes.\cite{Barbara:88,Reynolds:96}

A fluctuation of the electric field $\delta
\mathbf{R}(\Gamma)=\mathbf{R}(\Gamma) -\mathbf{R}_0$ out of
equilibrium requires reversible, non-expansion work applied to the
system;\cite{Landau5} this work is quantified by the Landau functional 
$F(\mathbf{R})$.  We will be interested here only in the projection of
the field on the direction $\mathbf{\hat m}_0=\mathbf{m}_0/ m_0$ of
the external dipole and thus will set up the ``reaction coordinate''
or ``order parameter'' $R= \mathbf{\hat m}_0 \cdot
\mathbf{R}$.\cite{Landau5} The free energy surface can then be found
by standard prescriptions\cite{ChaikinLubensky} singling out the order
parameter from the manifold of the system degrees of freedom
\begin{equation}
  \label{eq:7}
  e^{-\beta F(R)} \propto \int \delta \left(R - R(\Gamma) \right) e^{-\beta H} d\Gamma . 
\end{equation}
In this equation, $H$ is the Hamiltonian of the solution involving a
solute carrying no dipole ($\mathbf{m}_0=0$), and $\beta = 1/
(k_{\text{B}}T)$ is the inverse temperature. The free energy $F(R)$
thus describes thermal fluctuations of the electric field in the
absence of the dipole $\mathbf{m}_0$.

\begin{figure}
  \centering
  \includegraphics*[width=8cm]{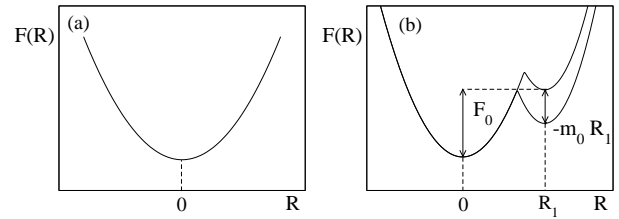}
  \caption{Sketch of the standard expectation (a) and the picture
    proposed in this study (b) for the free energy of creating field
    $R$ at a point within the solute when no dipole is present [Eq.\
    \eqref{eq:7}]. In our simulations, a point dipole is placed at
    that point, which modifies the free energy to $\mathcal{F}(m_0,R)$
    [Eq.\ \eqref{eq:9}]. The traditional linear-response approximation
    anticipates $F(R)$ to be harmonic in the whole range of $R$-values
    of interest. In contrast, we suggest the existence of an
    excitation state, with the corresponding minimum lifted by the
    free energy $F_0$. This excited state, too high to be observed in
    the homogeneous liquid or in non-polar solution, can become
    observable by placing the dipole $m_0$ into the solute, which
    lowers the free energy gap between the lower and higher minima by
    $ -m_0R_1$. The curvature of the excited state near its minimum
    might be different from the corresponding curvature at the lower
    minimum, thus altering the basic assumption of the linear-response
    approximation, the invariance of the spectrum of the solvent
    fluctuations with respect to the magnitude of $m_0$. The overall
    shift of $F(R)$ in the external field is not shown in the diagram;
    the positions of the lower minimum at zero and nonzero dipoles
    $m_0$ are therefore made coincide.  }
  \label{fig:1}
\end{figure}

Empirical observations suggest that the Onsager reaction field $R_0$
is a linear function of the dipole moment inducing
it.\cite{Barbara:88,Reynolds:96} This observation can be
mathematically cast into the requirement of a harmonic form of $F(R)$
(Fig.\ \ref{fig:1}a)
\begin{equation}
  \label{eq:8}
  F(R) = R^2/(2\kappa) . 
\end{equation}
The free energy changes when the dipole moment is turned on into
\begin{equation}
  \label{eq:9}
  \mathcal{F}(m_0,R) = - m_0 R + F(R) .
\end{equation}
The minimization of this free energy in respect to the field $R$
yields the reaction field linear in the solute dipole, $R_0 = \kappa
m_0$. This is the familiar linear response
approximation.\cite{Hansen:03} It articulates two physically
significant concepts: (i) the response of the medium (here electric
field) grows linearly with an external perturbation (solute dipole
moment) and (ii) the perturbation does not alter the spectrum of the
medium fluctuations which therefore can be calculated or measured from
the properties of the system in the absence of the perturbation
(zero-dipole solute in our case).

The distribution of the solvent electric field $P(R) \propto
\exp[-\beta F(R)]$ is a Gaussian function with the variance
$\sigma_R^2=\langle(\delta R)^2\rangle$ equal to $\kappa / \beta $,
$\delta R =R - R_0$.  Since fluctuations of the electric field
introduce fluctuations in the optical transition
energies,\cite{Pratt:94,DMjpca:01,Rajamani:04,Ben-Amotz:05} the stiffness parameter $\kappa$ (also known
as response function) is experimentally accessible from the
inhomogeneous width of an optical line. At the same time, since the
average energy, $e_0= - \kappa m_0^2$, and the distribution width are
given in terms of the same parameter $\kappa$, one gets the
fluctuation-dissipation\cite{Hansen:03} relation between the average
energy and the energy variance, $e_0 = - \beta \langle (\delta
e)^2\rangle$. The parameter
\begin{equation}
  \label{eq:10}
  \chi_G = -  \beta \langle (\delta e)^2\rangle/ e_0 
\end{equation}
can therefore be considered as the non-linearity parameter
quantifying deviations from the Gaussian statistics of the electric
field fluctuations.\cite{DMjpca:02} $\chi_G=1$ for the Gaussian
statistics when linear response holds and the stiffness parameter can
be calculated either from the reaction field or from the field
variance
\begin{equation}
  \label{eq:3}
 \kappa =  R_0/m_0 =  \beta \sigma_R^2 . 
\end{equation}

The inverse stiffness parameter $\kappa^{-1}$ has the meaning of the
characteristic volume in which a small probe dipole will be influenced
by the fluctuations of the electric field $R$. One can then introduce
the characteristic length to which the field fluctuations penetrate
inside the solute
\begin{equation}
  \label{eq:4}
  \ell_R = (\kappa)^{-1/3}
\end{equation}
This length will then define the thickness of the surface layer within
which a dipolar probe (e.g., optical dye) will be affected by the
solvent (i.e., the interaction energy will be greater than
$k_{\text{B}}T$).

The harmonic free energy profile [Eq.\ (\ref{eq:8})] sketched in Fig.\
\ref{fig:1}a is not the only concevable form of the free energy
surface. The reaction field $R_0$ building up in response to a growing
dipole moment $m_0$ is produced by the solvent dipoles increasingly
aligned with the external field. This externally forced alignment
leads to frustration of the local orientational structure within the
liquid, which can release itself into a new structure represented by a
second minimum of the Landau functional $F(R)$ (Fig.\ \ref{fig:1}b).

Several fundamental consequences might be anticipated if the situation
sketched in Fig.\ \ref{fig:1}b is realized. First, the curvature of
the free energy at the higher-energy minimum is not necessarily equal
to the curvature at the lower-energy minimum. In other words, the
fluctuation spectrum around the minimum $R_1$ might be altered
compared to the spectrum in the absence of the external field. If this
happens, the assumption (ii) listed above (no alteration to the
fluctuation spectrum) does not apply anymore. Further, the free energy
$F_0$ at the second minimum should be lifted compared to $F(0)=0$ in
expectation of no spontaneous field in the ground state.  However, the
field of the dipole $\mathbf{m}_0$ will bring the minimum of
$\mathcal{F}(m_0,R)$ down by $-m_0 R_1$ (Fig.\ \ref{fig:1}).  This
external-field effect offers a unique opportunity to study excitation
states of a liquid interface not significantly populated by thermal
excitations due to a high value of $F_0$.

Interfaces of water with solutes of nanometer dimension is a
particularly suitable candidate for observing surface excited
states. Large solutes break the network of hydrogen bonds of bulk
water producing unsatisfied surface bonds. The result is a specific
orientational structure of the surface waters with their dipoles
preferentially oriented parallel to the
interface.\cite{Lee:84,Sokhan:97,Rossky:10} While strong hydrogen-bond
interactions provide a sufficient driving force for in-plane
orientations, the electric field of external multipoles will compete
with this alignment, possibly switching it, in a discontinuous fashion,
into a new interfacial structure. In fact, Sharp and co-workers have
found that water at the solute interface occupies two states of local
structure characterized by straight, ``ice-like'' hydrogen bonds
around non-polar solutes and bent bonds, required by the dipoles
alignment along the electrostatic field of polar/ionic
solutes.\cite{Sharp:97,Sharp:10} The distinction between polar and
non-polar solvation can be attributed in this picture to changes in
occupations of these two hydrogen bonding states. What has been still
missing from that picture is the ability of surface waters to switch
between the two states via a concerted cross-over. The existence of
such a cross-over, occurring with increasing electrostatic field of
the solute, is the main finding of this study.

\begin{figure}
  \centering
  \includegraphics*[width=6cm]{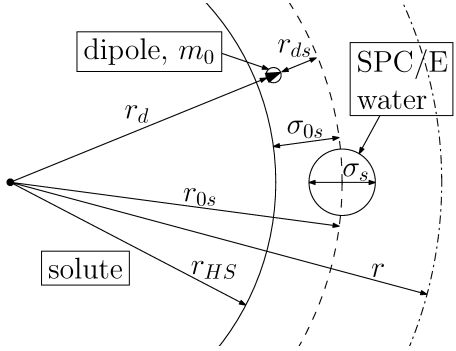}
  \caption{Cartoon of the solute-solvent configuration.  The solute is
    a hard core of radius $r_{\text{HS}}$ covered with the LJ layer of
    the width $\sigma_{0s}$ [Eq.\ \eqref{eq:2}]. The distance
    $r_{0s}=r_{\text{HS}}+\sigma_{0s}$ approximately corresponds to
    the first peak of the solute-solvent pair distribution function
    $g_{0s}(r)$ [also see Fig.\ \ref{fig:8} below]. The point dipole
    $m_0$ is placed at the distance $r_d$ from the solute center and,
    correspondingly, the distance $r_{ds}=r_{0s}-r_{\text{HS}}$ from
    the solute-solvent interface. The solvent is SPC/E water at 1 atm
    and 273 K (the melting temperature of SPC/E water is 215
    K\cite{vega:114507}). }
  \label{fig:2}
\end{figure}

\begin{figure}
  \centering
  \includegraphics*[width=6cm]{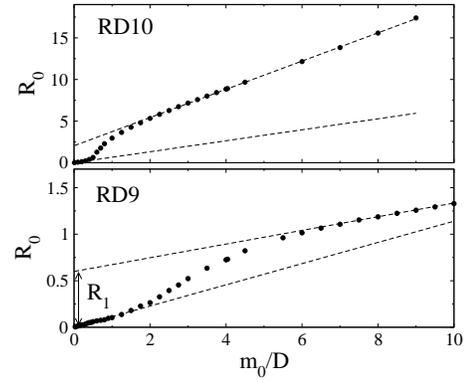} 
  \caption{Reaction field $R_0$ (in V/\AA) vs solute dipole $m_0$ for
    RD9 ($r_d=9$ \AA) and RD10 ($r_d=10$ \AA) configurations (see
    Fig.\ \ref{fig:2}). The dashed lines show the slopes of the linear
    portions of $R_0(m_0)$. The scaling of $R_0$ with $m_0$ is
    approximately quadratic in the intermediate region. The vertical
    arrow in the lower panel indicates the equilibrium field $R_1$ in
    the excited state obtained by extrapolating the reaction field in
    the excited configuration to zero solute dipole (Fig.\
    \ref{fig:1}). }
  \label{fig:3}
\end{figure}

The system we study here is a non-polar solute immersed in SPC/E
water\cite{vega:114507} and interacting with it by a Kihara
potential characterized by a hard-sphere core, $r_{\text{HS}}=9$ \AA, and
two Lennard-Jones (LJ) parameters, $\sigma_{0s}=3$ \AA\ and
$\epsilon_{0s} = 0.65$ kJ/mol, 
\begin{equation}
  \label{eq:2}
  \phi_{\text{0s}}(r) = 4\epsilon_{0s}\left[\left(\frac{\sigma_{\text{0s}}}{r-r_{\text{HS}}}\right)^{12} 
            - \left(\frac{\sigma_{\text{0s}}}{r-r_{\text{HS}}}\right)^6\right].
\end{equation} 
We then modify the system by introducing a dipole moment at the
distance $r_d$ from the solute center (Fig.\ \ref{fig:2}) and watch the
changes in the distribution of the water field $R$ experienced by the
dipole when the dipole magnitude grows.  While the system follows the
standard expectations of linear response at low solute dipoles $m_0$,
the situation changes at some critical value of the solute dipole when
a new, non-linear, functional form for the reaction field emerges
(Fig.\ \ref{fig:3}). The reaction field returns to a linear trend
with $m_0$ when the dipole keeps growing, but extrapolates to a
non-zero value at $m_0\rightarrow 0$, as would be expected if the
system has landed in a high free-energy minimum, as in Fig.\
\ref{fig:1}b. Further, the curvature around this second minimum (slope
of $R_0$ vs $m_0$ in Fig.\ \ref{fig:3}) also experiences alteration,
as we have described above. The basic physical picture is the
emergence of a finite domain of surface waters in an excited state
different both from the bulk and from the rest of the interface. The
details of the simulation protocol can be found in the Supplementary
Material (SM)\cite{supplJCP} and we proceed to the discussion of the
results.

\section{Three-state model}
Before proceeding to a more detailed description of the simulation 
results, we briefly outline the analytical framework used here to
understand the outcome of the simulations. This is a phenomenological
three-state model representing the simulation data in terms of three
distinct states of surface waters. 

The model includes three states described by Landau free energy
functions $F_{\alpha}(R)$ [Eq.\ (\ref{eq:7})] with $\alpha=g,1,2$
standing for the ground state (g) and two exited states corresponding
to the minima $R_{\alpha}$ at the positive (1) and negative (2)
reaction field values.  Two excited states, instead of one as in Fig.\
\ref{fig:1}, are introduced to satisfy the condition of vanishing
reaction field at $m_0\rightarrow 0$.  The minima of these states are
lifted by the energies $F_{\alpha}$ from the free energy minimum
$F_g(0)$ of the ground state. Each of $F_{\alpha}(R)$ curves is
modified by the presence of an external dipole $m_0$ into
$\mathcal{F}_{\alpha}(m_0,R) = F_{\alpha}(R) - m_0 R$, as in Eq.\
(\ref{eq:9}). The global free energy of the system is then found as
the statistical trace over the three states as follows
\begin{equation}
  \label{eq:11}
  e^{-\beta \mathcal{F}(m_0,R)} = \sum_{\alpha} e^{-\beta \mathcal{F}_{\alpha}(m_0,R)} . 
\end{equation}

From MD simulations, we have found little asymmetry in the
distribution of water's dipolar directions at the interface, as judged
from the first-order orientational parameter
\begin{equation}
  \label{eq:12}
  p_1^I = (N^I)^{-1} \sum_{j=1,N^I} \mathbf{\hat m}_j \cdot \mathbf{\hat r}_j .
\end{equation}
Here, the sum runs over $N^I$ waters in the solute's first hydration
layer, $\mathbf{\hat m}_j = \mathbf{m}_j/m$, and $\mathbf{\hat r}_j =
\mathbf{r}_j/r_j$ is the unit vector defining the position of the
first-shell water relative to the solute's center.  The order
parameter $ p_1^I$ is close to zero indicating that dipoles oriented
parallel and antiparallel to the surface normal occur with
approximately equal probability. Therefore, the average field at
$m_0=0$
\begin{equation}
  \label{eq:15}
  \langle \mathbf{R} \rangle = \sum_j \mathbf{T}(\mathbf{r}_d -\mathbf{r}_j) \cdot
  \langle \mathbf{m}_j \rangle,
\end{equation}
($\mathbf{T}$ is the dipolar tensor) is expected to approach zero. 

One needs to mention that the orientational structure of surface water
around spherical cavities and non-polar solutes results in a positive
value of the cavity electrostatic
potential.\cite{Ashbaugh:00,Rajamani:04} The potential is apparently
rather uniform resulting in $\langle \mathbf{R} \rangle \simeq 0$ in both the
present simulations with the off-center dipole, as well as in our
previous study of the same Kihara solute with the dipole placed at the
solute's center.\cite{DMcpl:11}

\begin{figure}[h]
  \centering
  \includegraphics*[width=6cm]{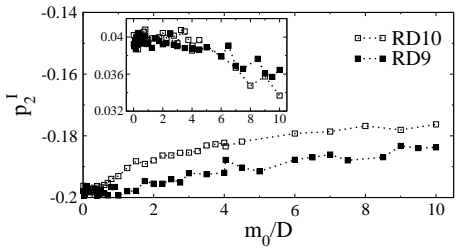}
  \caption{Second [$p_2^I$, Eq.\ \eqref{eq:13}] (main panel) and first
    [$p_1^I$, Eq.\ \eqref{eq:12}] (inset) orientational order
    parameters for RD9 ($r_d=9$ \AA, filled squares) and RD10
    ($r_d=10$ \AA, open squares) configurations. }
  \label{fig:4}
\end{figure}

In contrast to the first-order orientational order parameter $p_1^I$ the second-order
orientational parameter 
\begin{equation}
  \label{eq:13}
  p_2^I =  (2N^I)^{-1} \sum_{j=1,N^I} 
\left[3(\mathbf{\hat m}_j \cdot \mathbf{\hat r}_j)^2 - 1 \right]
\end{equation}
is of the order $p_2^I \simeq -0.19$, indicating that about 40\% of
first-shell waters have their dipoles directed perpendicular to the
surface normal.\cite{Lee:84,Sokhan:97,Rossky:10} This parameter decays
only slightly upon the insertion of the solute dipole (Fig.\
\ref{fig:4}).

Taken together, the results for $p_{1,2}^I$ suggest that there is
little asymmetry in the distribution of the interfacial dipoles.  The
condition $\langle R\rangle \simeq 0$ in Eq.\ \eqref{eq:15} then
implies symmetric $\mathcal{F}(0,R)$, which can be achieved with a
reduced number of model parameters: $R_{2}=-R_{1}$ and
$F_{1}=F_{2}=F_0$. The free energies of the three states then become
\begin{equation}
  \label{eq:14}
  \begin{split}
    F_g(R) = & R^2 / (2\kappa_g) \\
    F_1(R) = & (R-R_1)^2 /(2 \kappa_1) + F_0\\
    F_2(R) = & (R+R_1)^2/ (2 \kappa_1) + F_0
  \end{split}
\end{equation}
Four model parameters $\kappa_g$, $\kappa_1$, $R_1$, and $F_0$ were
fitted to $\sigma_R^2(m_0)$ from MD data.  They are listed in Table
\ref{tab:1} and are used to produce plots of $R_0(m_0)$,
$\sigma_R^2(m_0)$, and $\mathcal{F}(m_0,R)$ presented in the Results
section below.

\begin{table}[h]
  \centering
  \caption{List of model parmeters produced by fitting the three-state model 
    to $\sigma^2_R(m_0)$ from MD
    simulations. $\ell_R^{\alpha}=(\kappa_{\alpha})^{-1/3}$, $\alpha
    = \mathrm{g,1}$ is the
    characteristic length of penetration of water's surface
    fluctuations into the solute [Eq.\ \eqref{eq:4}]. }
  \label{tab:1}
  \begin{tabular}{lccccc}
 \hline
 Configuration & $r_d$, \AA  & $\beta F_0$ &  $\ell_R^g$, \AA & $\ell_R^1$, \AA
 & $R_1$, V/\AA \\
 \hline
  RD9  & 9  & 7.3 & 3.0 & 3.1  & 0.3 \\
  RD10 & 10 & 7.2 & 1.5 & 1.2  & 1.0 \\
 \hline
  \end{tabular}
\end{table}

\section{Results}
All the simulation results that we discuss below have been produced
for two distances between the point dipole in the solute and its
center (Fig.\ \ref{fig:2}), $r_d=9$ \AA\ and $r_d=10$ \AA. We will
reference them throughout below as states RD9 and RD10, respectively.

\begin{figure}
  \centering
  \includegraphics*[width=6cm]{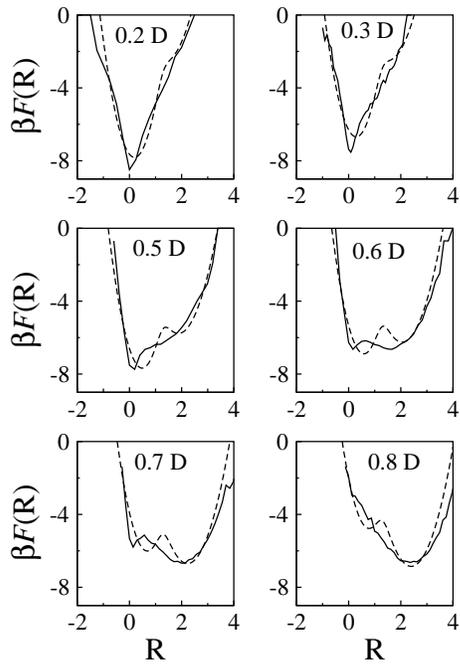}
  \caption{Free energies $\mathcal{F}(m_0,R)$ [Eq.\
    \eqref{eq:9}] for the RD10 state ($r_d=10$ \AA) at $m_0$ values
    indicated in the plots. The reaction
    field in the abscissa is in V/\AA; the dashed lines refer to the
    fits to the three-state model [Eqs.\ \eqref{eq:11}, \eqref{eq:14}].
    The model parameters used in the plot are listed in Table \ref{tab:1}.  }
  \label{fig:5}
\end{figure}

There is a fairly broad intermediate region between two linear scaling
regimes of the reaction field $R_0$ with the solute dipole $m_0$. This
intermediate, approximately quadratic, scaling is a signature of
transitions between the low and high free energy states. This is
clearly seen in the progression of free energies $\mathcal{F}(m_0,R)$
with changing $m_0$ shown in Fig.\ \ref{fig:5} for the RD10 state. The
solid lines in the figure are obtained directly from distributions of
$R$ from MD simulations, while the dashed lines are the fits to the
three-state model outlined above. The overall picture of the
alteration of the free energy surface with the external dipole is
remarkably consistent with the sketch shown in Fig.\ \ref{fig:1}b
pointing to a discontinuous, first-order character of the cross-over
between the ground and excited surface states.

\begin{figure}[h]
  \centering
  \includegraphics*[width=6cm]{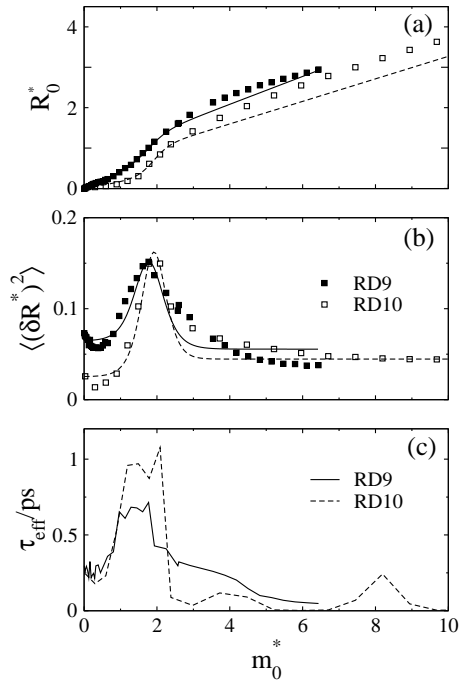}
  \caption{The average, $R_0^*$, of the reduced field $R^*$ as defined
    by Eq.\ \eqref{eq:6} (a) and its variance (b) vs the reduced
    solute dipole defined by Eq.\ \eqref{eq:1}. The filled squares and
    solid lines in (a) and (b) refer, respectively, to the MD results
    and their fit to the three-state model for the RD9 configuration;
    the open squares and the dashed lines carry the same information
    for the RD10 configuration. Panel (c) shows the relaxation time
    $\tau_{\text{eff}}$ of the water field self-correlation function
    (see text) obtained from MD simulations for RD9 (solid line) and
    RD10 (dashed line) configurations. }
  \label{fig:6}
\end{figure}

As is expected from the general phenomenology of discontinuous
transitions,\cite{Landau5,Blinc:74,Binder:92} the fluctuations of the
order parameter are non-Gaussian close to the transition point,
reflecting transitions of the system between the two alternative
states. Such enhanced fluctuations are typically reflected by spikes
of variances, such as the heat capacity.\cite{Binder:92} Our
observations are no different and the variance $\sigma_R^2 = \langle
(\delta R)^2 \rangle$ shows a peak at the transition value of the
solute dipole moment (Fig.\ \ref{fig:6}b). The results are plotted
against a dimensionless dipole magnitude that reflects the ratio of
the characteristic field produced by the solute to the one existing
within the solvent
\begin{equation}
  \label{eq:1}
    m_0^* = \frac{m_0}{m_s} \left(\frac{\sigma_s}{r_{ds}}\right)^3 ,
\end{equation}
where $m_s$ and $\sigma_s$ are the solvent (water) dipole moment and
diameter, respectively. In addition, $r_{ds}=r_{0s}-r_{\text{HS}}$
(Fig.\ \ref{fig:2}) is the distance between the point dipole and the
solute-water interface. Similar ideas are used to introduce the
reduced instantaneous field 
\begin{equation}
  \label{eq:6}
  R^*=  R r_{ds}^3/m_s 
\end{equation}
the average, $R_0^*$, and variance, $\langle(\delta R^*)^2\rangle$, of
which are shown in Fig.\ \ref{fig:6}.

The spike in the field variance is accompanied by a strong increase in
the relaxation time of the normalized time self-correlation function,
$S(t)=\langle \delta R(t) \delta R(0) \rangle/ \langle (\delta R)^2
\rangle$. The relaxation time shown in Fig.\ \ref{fig:6}c was
calculated as $\tau_{\text{eff}}=\int_0^{\infty} dt S(t)$. The
appearance of the peak in the relaxation time is reminiscent of
critical slowing down well established for phase
transitions,\cite{Blinc:74,Binder:92} but how far that language can be
carried over to our situation is not entirely clear.

\begin{figure}[h]
  \centering
  \includegraphics*[width=6cm]{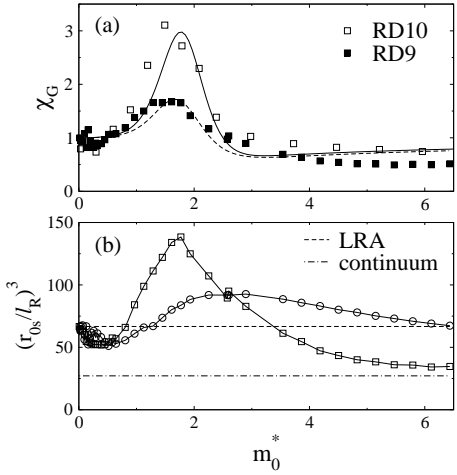}
  \caption{Nonlinearity parameter $\chi_G$ [Eq.\ \eqref{eq:10}] (a)
    obtained from MD simulations (points) and from fitting the
    three-state model (lines). Panel (b) shows the reduced stiffness
    parameter $\kappa r_{0s}^3=(r_{0s}/ \ell_R)^3$ calculated in
    different approximations for RD9. The dash-dotted line, 
    $\kappa=R_0/m_0$, is from the dielectric continuum calculation 
    for the reaction field at a dipole inside a dielectric cavity [Eq.\
    \eqref{eq:5}]. The dashed line shows $\kappa=\beta\sigma_R^2$
    obtained from the variance $\sigma_R^2$ at $m_0=0$. The open
    circles show $\kappa=R_0/m_0$ and open squares refer to
    $\kappa=\beta \sigma_R^2$, both from MD simulations. In linear
    response, circles and squares are expected to collapse on the
    dashed line, Eq.\ \eqref{eq:3}.  }
  \label{fig:7}
\end{figure}

\begin{figure}[h]
  \centering
  \includegraphics*[width=6cm]{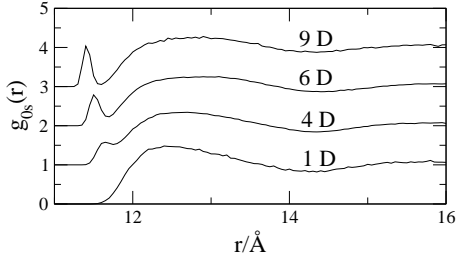}
  \caption{Solute-oxygen pair correlation function for RD9, inside 
    the solid angle defined in the text (see also the 
    SM\cite{supplJCP}). The numbers in the plot indicate values of 
    the dipole moment $m_0$. The curves are shifted vertically by 
    1.0 for a better view. }
  \label{fig:8}
\end{figure}

The preceding discussion clearly indicates that fluctuations of
water's electric field are expected to be non-Gaussian in the range of
solute dipoles allowing switching of surface waters between the ground
and excited states in the spirit of Figs.\ \ref{fig:1} and
\ref{fig:5}. We indeed observe a spike in the nonlinearity parameter
$\chi_G$ [Eq.\ \eqref{eq:10}] in the transition region (Fig.\
\ref{fig:7}a). The extent and the breadth of this non-Gaussian
behavior are additionally illustrated in Fig.\ \ref{fig:7}b where the
direct calculation of the stiffness parameter $\kappa$ from the
reaction field, $\kappa= R_0/ m_0$, (circles) is compared to the
expectation from the linear response according to Eq.\ \eqref{eq:3},
$\kappa=\beta \sigma_R^2$ (squares). Since the linear response
function is independent of the dipole moment, the horizontal dashed
line in Fig.\ \ref{fig:7}b indicates the value to which both
calculations would converge if linear response held. In order to
provide an additional reference point to our results, the horizontal
dash-dotted line shows $\kappa$ for a point dipole placed inside a
cavity in dielectric continuum. The reaction field in this
approximation is derived in the SM\cite{supplJCP}
\begin{equation}
  \label{eq:5}
  R_0 = \frac{m_0}{ r_{0s}^3} \frac{\epsilon-1}{2 \epsilon + 1 } 
   \left(2-\frac{z^2}{\epsilon+1} \right)
   \frac{1+z^2}{\left( 1-z^2 \right)^3},  
\end{equation}
where $z=r_d/r_{0s}$.  The continuum cavity radius in this calculation
was chosen at $r_{0s}=11.5$ \AA, slightly below the position of the
first peak of the solute-solvent pair distribution function at $\simeq
12$ \AA (Fig.\ \ref{fig:8}).

The results presented here pose a significant question of the nature
of the excited state of the surface water. A related question is what
is the spatial extent of the excited water domain. The results for the
alteration of the orientational order parameters with changing $m_0$ (
Fig.\ \ref{fig:4}) support the notion of a relatively localized
perturbation, not spreading into the entire hydration layer.

In order to address the question of the water structure in a domain
nearest to the solute dipole we have separated a solid angle of
27.4$^o$ drawn with the dipole direction as the axis and cutting a
portion of the surface circumference equal to $2 \sigma_s$ (see
SM\cite{supplJCP} for a drawing). The pair solute-water distribution
function was calculated for waters residing within this solid angle only
and is shown in Fig.\ \ref{fig:8}.  A new peak emerges from the
nearest solute-water distance with increasing $m_0$. The area under
the peak shows that exactly one water molecule breaks from the surface
hydrogen-bond network, shifting toward the solute. This interpretation
is consistent with the excitation free energy of $F_0\simeq 7$
$k_{\text{B}}T$ (Table \ref{tab:1}), close to the enthalpy required to
break two hydrogen bonds in SPC/E water.\cite{Spoel:2006kx}

\begin{figure}
  \centering
  \includegraphics*[width=6cm]{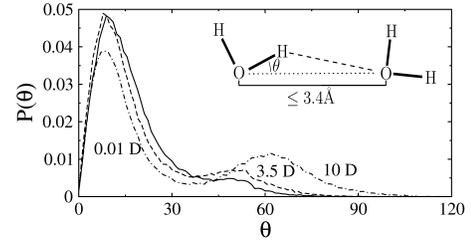}
  \caption{Distributions of O-H-O angles $\theta$ for the RD10
    configuration at $m_0$ equal to 0.01, 3.5, and 10 D.  The
    distributions were calculated according to the algorithm suggested
    by Sharp \textit{et al}\cite{Sharp:97} in the region of the first
    hydration layer within the solid angle cutting the circumference
    length of $2\sigma_s$ at the surface (see also the
    SM\cite{supplJCP}). The inset shows the definition of the bond
    angle and the distance cutoff within which the angles were
    sampled. }
  \label{fig:9}
\end{figure}

The creation of the excitation defect in the hydration shell results
in the frustration of the nearest bonds network seen in the alteration
of the distribution of O-H-O angles of first-shell waters within the
same solid angle as used in producing Fig.\ \ref{fig:8}.  These
distributions at changing $m_0$ are shown in Fig.\ \ref{fig:9} Similar
to the previous observations by Sharp and co-workers,\cite{Sharp:97}
the population of the buckled, large-angle state grows with increasing
$m_0$. This is a behavior generic for polar (ionic or dipolar) solutes
frustrating water's hydrogen-bond network by their electric fields.

\section{Discussion}
We have studied the statistics of electric field fluctuations produced
by surface water inside the solute. The main finding of this study is
the realization that the statistics can be profoundly altered by
placing a dipole next to the interface. With increasing dipole
magnitude, waters closest to the dipole alter their
hydrogen-bond network in a way that creates a domain of surface waters 
in an excited state, with the free energy lifted by approximately the
energy of breaking two hydrogen bonds ($\simeq 7$ $k_{\text{B}}T$,
Table \ref{tab:1}). A crude estimate of what is needed to shift the
system into this excited state in provided by Fig.\ \ref{fig:6}b: a
dipole twice as large as water's dipole placed one water diameter from
the interface puts the system at the cross-over point at the peak of
non-Gaussian fluctuations. This outcome seems to be quite generic:
once surface particles have close probabilities to occupy alternative
states, the variance of the fluctuations is in excess to the
expectation from the linear response.\cite{DMcp:08}

The local nature of the excitation, most likely limited to one water
breaking from the network and creating a corresponding elastic
deformation around it, contributes to a significant range of dipole
moment magnitudes at which the system finds itself distributed between
two local minima, each characterized by a linear response.  This
property of the local surface transition, clearly distinct from much
more localized, in the parameters space, phase
transitions,\cite{Binder:92} will allow one to observe the effects
discussed here for a substantial range of solute perturbations.

The surface defect and the corresponding frustration of the network
characterizing the excitation do not necessarily require a surface
dipole. They might be achieved by other causes such as ionic or
hydrogen-bonding surface sites. An optical probe placed near such a
domain of waters in the excited state will record an optical response
distinct in both the spectral shift and the width from probes next to
the ground-state interface. In this regard, the inhomogeneous spectral
broadening, which is admittedly hard to extract experimentally, is
nevertheless a more sensitive probe of the local interfacial structure
than the spectral shift. The Stokes shift dynamics can also slow down
in the transition region, as is seen in Fig.\ \ref{fig:6}c. A mosaic
of ground- and excited-states domains at an extended interface of a
nanometer-scale solute will then contribute to energetic and dynamic
heterogeneity of the electric field
fluctuations.\cite{Giovambattista:08} However, the dynamics of the
electric field observed here (ps) are still significantly faster than
that found for hydrated proteins
(ns).\cite{Abbyad:07,Tripathy:10,DMjpcb1:11} The three orders of
magnitude difference suggests a possibility of cooperativity between
many surface excitations at proteins' polar/ionic surface
sites. Alternatively, the excited surface state of a protein might
still be localized, and the slow dynamics come as a result of protein
motions modulating the interfacial polarization.\cite{Halle:09}

\acknowledgments This research was supported by the National Science
Foundation (CHE-0910905). CPU time was provided by the National
Science Foundation through TeraGrid resources (TG-MCB080116N).

\bibliography{chem_abbr,dielectric,dm,statmech,protein,liquids,solvation,dynamics,glass,elastic,et,surface}

\end{document}